\documentclass[twocolumn,secnumarabic,amssymb, nobibnotes, aps, prl]{revtex4-2}

\usepackage{graphicx}
\usepackage{dcolumn}
\usepackage{bm}
\usepackage{amsmath}
\usepackage{lineno}
\usepackage[colorlinks=true,linkcolor=blue,citecolor=blue,urlcolor=blue]{hyperref}

\begin{document}
\preprint{APS/123-QED}

\title{\textbf{Observation of Entanglement Enabled Spin-Interference in the Drell-S${\rm \ddot{o}}$ding Process in Au+Au Ultraperipheral Collisions at RHIC} 
}%

\collaboration{The STAR Collaboration}
\noaffiliation




\begin{abstract}
We report a measurement of the Drell-S${\rm \ddot{o}ding}$ $\pi^+ \pi^-$ production in Au $+$ Au ultraperipheral collisions at center-of-mass energy per nucleon pair $\sqrt{s_{NN}} = 200$ GeV using the STAR detector. For the first time, the Entanglement Enabled Spin-Interference (EESI) effect is observed in the Drell-S${\rm \ddot{o}ding}$ process through the amplitude ${\rm A_{2\Delta\phi}}$ of the ${\rm cos(2\Delta\phi)}$ modulation. 
The measured ${\rm A_{2\Delta\phi}}$ exhibits no significant dependence on the invariant mass $M_{\pi^+ \pi^-}$. An enhancement of three standard deviations is observed for the Drell-S${\rm \ddot{o}ding}$ process compared to the $\rho^0$ photoproduction.
Furthermore, we measure the $|t|$-dependence of the Drell–S${\rm \ddot{o}ding}$ process and $\rho^0$ photoproduction. The Drell–S${\rm \ddot{o}ding}$ spectrum falls more steeply with $|t|$ and exhibits diffractive structures shifted to lower $|t|$ compared to $\rho^0$ photoproduction, features that are not captured by theoretical calculations.
These results provide new insights into the interplay between quantum interference and photon-nuclear interactions in ultra-peripheral heavy-ion collisions.

\end{abstract}

\maketitle

Ultra-strong electromagnetic fields in relativistic heavy-ion collisions are manifested as a comoving photon cloud accompanying the nuclei~\cite{weizsacker1934ausstrahlung,williams1934nature}, enabling photonuclear interactions. In these processes, a photon fluctuates into a virtual vector meson or hadron-antihadron pair, which scatters via Pomeron exchange (gluonic exchanges). Such interactions provide a unique probe of the photon's hadronic structure~\cite{bauer1978hadronic,AbdulKhalek:2021gbh,Li:2022kwn} and serve as a complementary tool for studying key nuclear properties, including nuclear gluon distributions~\cite{guzey2013exclusive} and the overall size of the nucleus~\cite{sciadv.abq3903,Klein:1999qj}.

A quasi-real spin-1 photon can fluctuate into (\romannumeral1) a virtual vector meson $(J^{PC}=1^{--})$, or (\romannumeral2) a virtual hadron-antihadron pair, giving rise to non-resonant photoproduction, known as the Drell-S${\rm \ddot{o}ding}$ process. 
In the case of scalar mesons such as $\pi^{+}\pi^{-}$, the angular momentum is encoded in the pair orbital momentum, resulting in quantum entanglement between the $\pi^{+}$ and $\pi^{-}$ in the relative-motion spatial wave function. A separate quantum phenomenon arises from the collision geometry: since both nuclei can act as photon emitters or targets, the emission source (nucleus A or B) becomes indistinguishable, establishing a ``double-slit'' interference pattern at the femtometer scale~\cite{sciadv.abq3903,PhysRevD.103.033007}. This interference induces second-order cosine oscillations in the asymmetries of the decay angular $\Delta\phi$ distributions~\cite{xing2020cos}. 
In this picture, interference at the pair level arises from the indistinguishability of the photon emission sources, while the entanglement of the $\pi^{+}\pi^{-}$ final state allows this interference to manifest in the measured decay angular distributions, giving rise to EESI effect.

\begin{figure}[t]
\centering

\begin{minipage}{0.325\columnwidth}
\centering
\includegraphics[width=\linewidth]{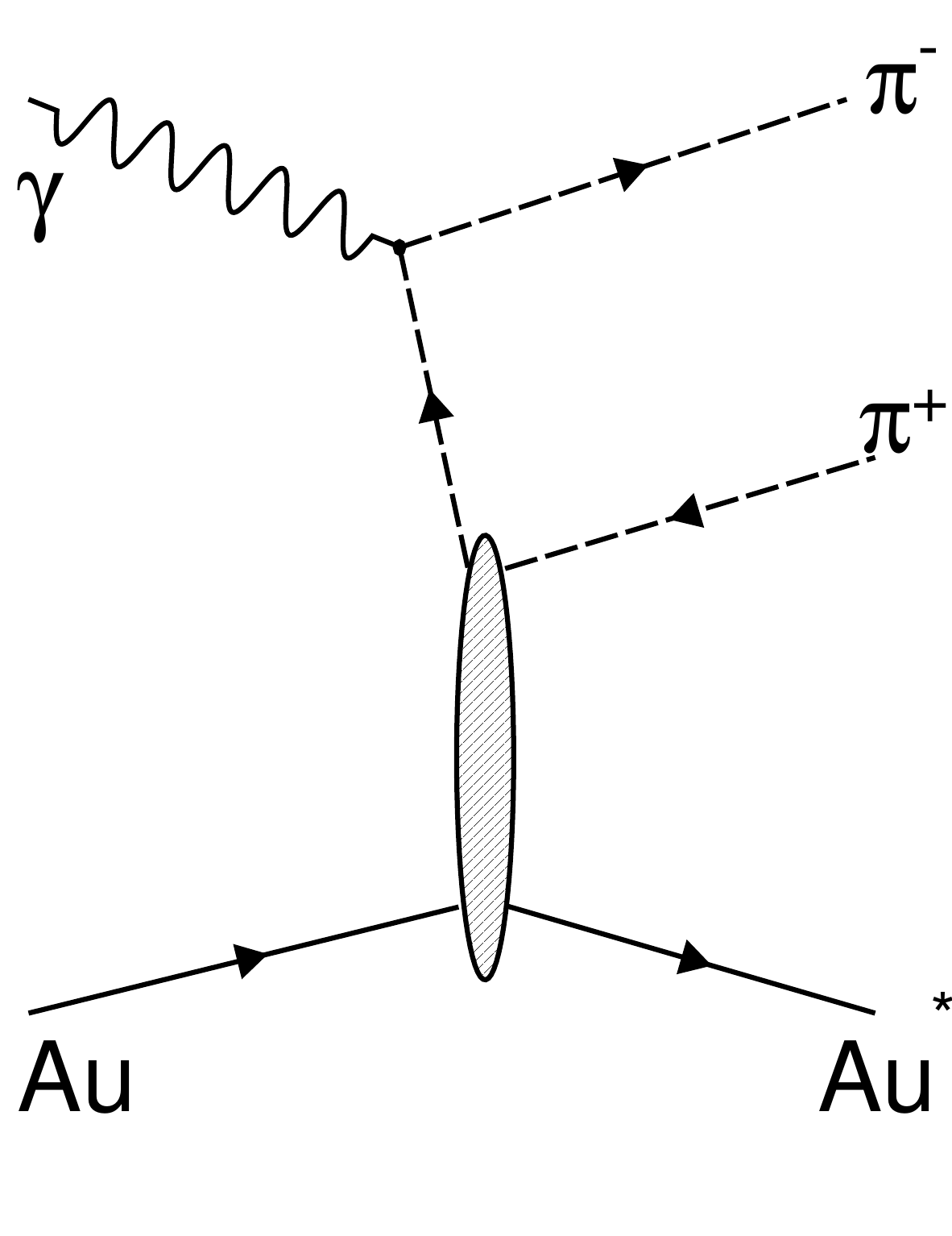}\hfill
\end{minipage}
\begin{minipage}{0.325\columnwidth}
\centering
\includegraphics[width=\linewidth]{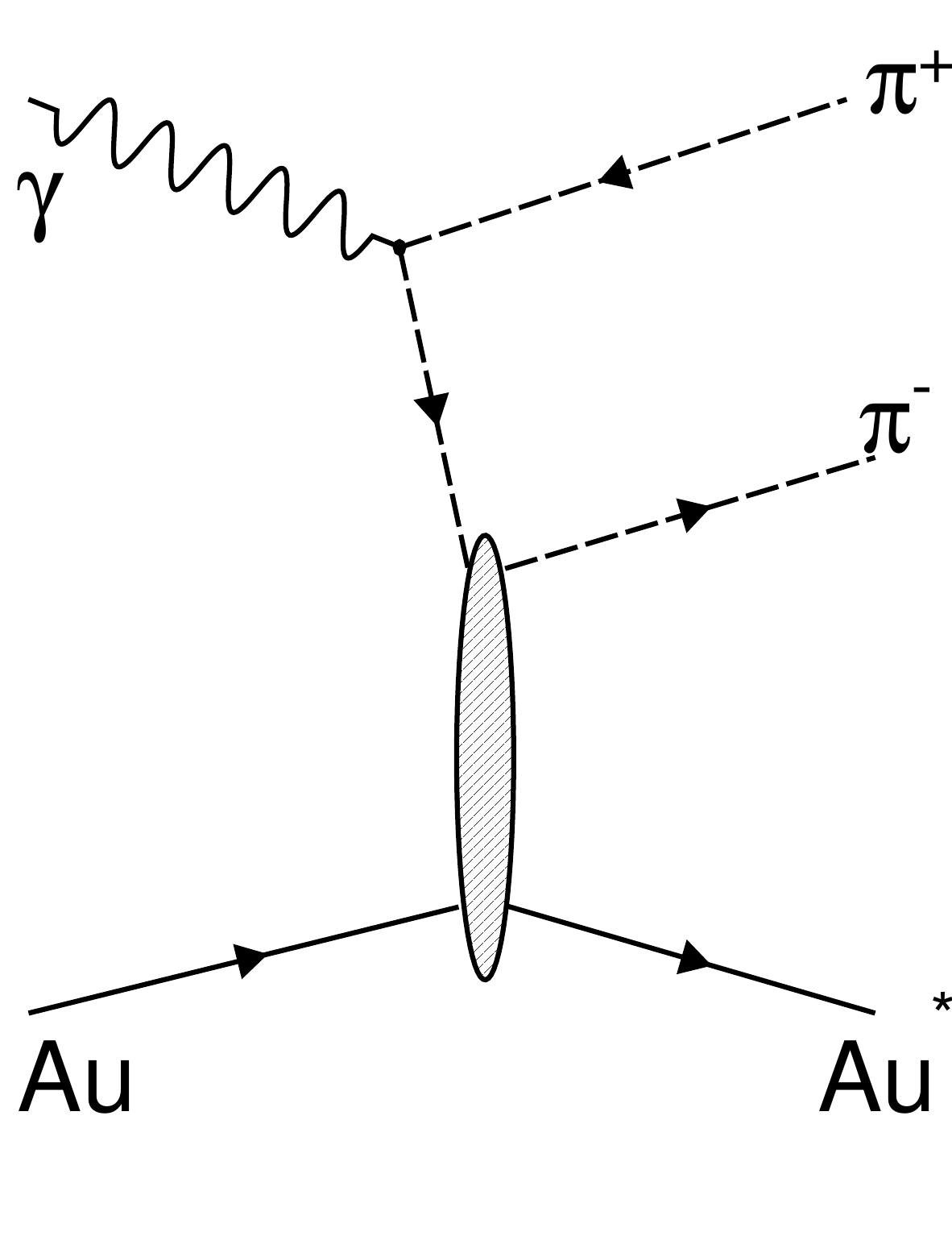}\hfill
\end{minipage}
\begin{minipage}{0.325\columnwidth}
\centering
\includegraphics[width=\linewidth]{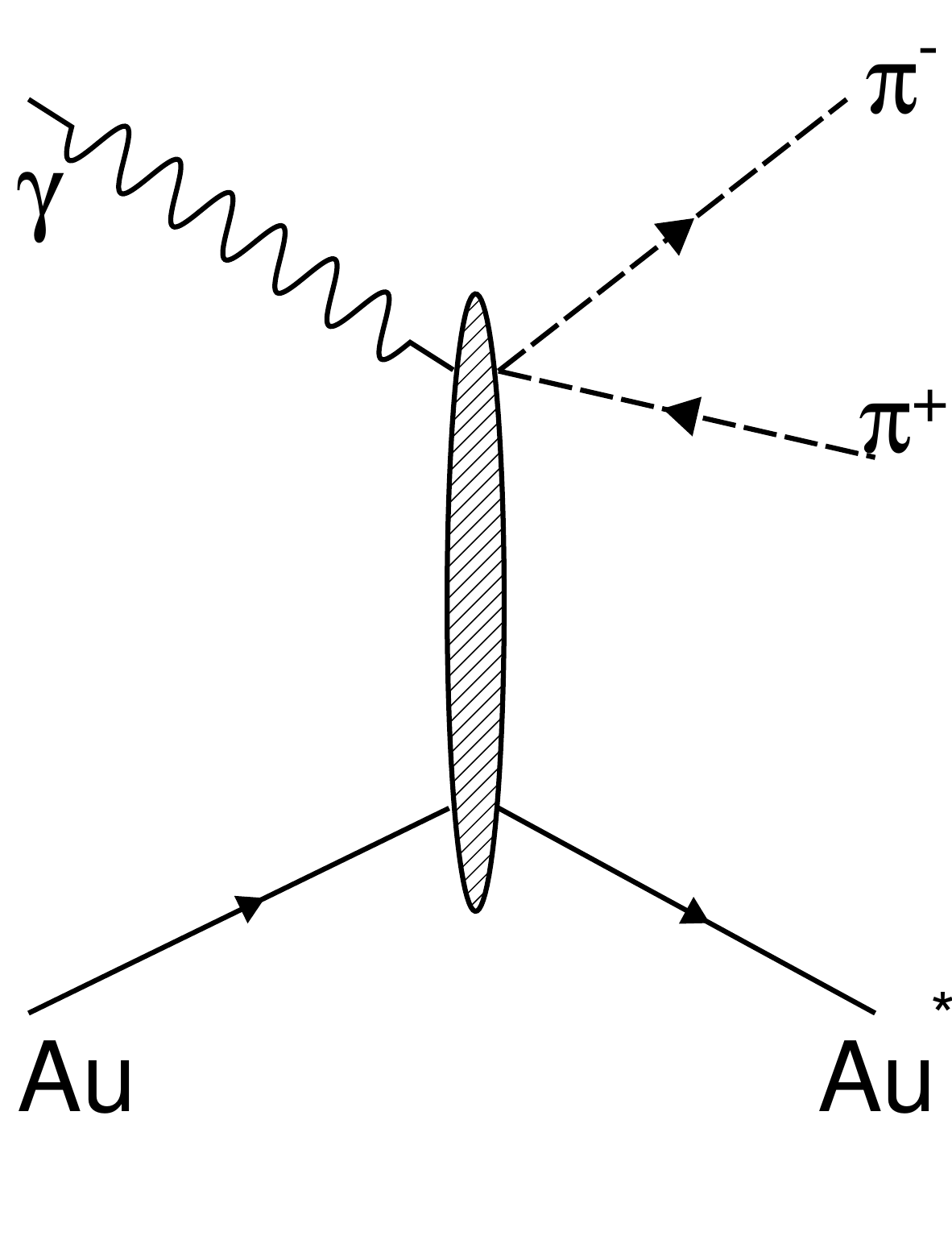}
\end{minipage}
\caption{Diagrams of Drell-S${\rm \ddot{o}ding}$ $\pi^{+}\pi^{-}$ production: (left and middle) dominant diagrams corresponding to $\pi^+-$nucleus and $\pi^--$nucleus scattering; (right) $\pi^+\pi^--$nucleus coupling diagram}
\label{fig:1}
\end{figure}

The Drell–S${\rm \ddot{o}ding}$ process is a sensitive probe of the nuclear gluon structure, nuclear gluon saturation, and shadowing effects~\cite{ZEUS:1997rof,Laget:2019tou}.
It has two dominant diagrams, corresponding to $\pi^+-$nucleus and $\pi^--$nucleus scattering~\cite{pumplin1970diffraction,bolz2015photoproduction}, as depicted in Fig.~\ref{fig:1}. A sub-leading $\pi^+\pi^--$nucleus coupling diagram is included to ensure gauge invariance\cite{pumplin1970diffraction,Wu:2026emu}.
In 1960, Drell first proposed these diagrams as promising contributors to secondary beams at SLAC~\cite{drell1960production}. Later in 1966, S${\rm \ddot{o}ding}$ carried out a full calculation of the $\gamma p \to \pi^+\pi^- p$ process and used it to explain the observed skewing of the $\rho^0$ resonance peak~\cite{soding1966apparent}. 
In diffractive photoproduction, the momentum-transfer $|t|$ distribution primarily reflects the transverse spatial profile of the target through its form factor. At the same time, it can also be sensitive to the transverse size of the interacting dipole via the convolution of the photon and hadronic wavefunctions with the scattering amplitude. In particular, a steeper $|t|$ dependence is generally associated with larger effective transverse configurations~\cite{Kuroda:2017ogq,Kowalski:2006hc,Nemchik:1996cw}.
In this context, comparing the $|t|$ spectra of the resonant and non-resonant $\pi^+\pi^-$ contributions within the same mass and rapidity range provides a way to probe potential differences in their underlying transverse dipole sizes.

Vector meson photoproduction, e.g., $\rho^0$ and $J/\psi$, has been measured extensively in ultraperipheral collisions (UPCs) of heavy ions~\cite{alice2021coherent,acharya2019coherent,STAR:2017enh,abelev2008rho,STAR:2019yox,STAR:2002caw,afanasiev2009photoproduction,guzey2013evidence,ALICE:2020ugp,ALICE:2024ife,collaboration2021first,GlueX:2024erj,Pybus:2024ifi,GlueX:2019mkq}. 
However, the non-resonant Drell-S${\rm \ddot{o}ding}$ process remains poorly characterized due to its small cross section and overlap with resonant $\rho^0$.
The Drell-S${\rm \ddot{o}ding}$ term cannot be isolated on an event-by-event basis and is extracted statistically after subtracting known contributions, relying on its interference with the $\rho^0$ photoproduction.
In previous $\rho^0$ measurements, the Drell-S${\rm \ddot{o}ding}$ component is treated with an unknown amplitude formulated as a constant~\cite{STAR:2017enh,abelev2008rho,STAR:2019yox,STAR:2002caw,ALICE:2020ugp} or a parameterised function~\cite{H1:2020lzc,ZEUS:2011tzw}. 
The extracted Drell-S${\rm \ddot{o}ding}$ contribution depends on the assumed mass dependence of this term. Therefore, adopting a more realistic spectrum derived from model calculations would be desirable to achieve a more precise measurements of the Drell-S${\rm \ddot{o}ding}$ process.
With growing interest in exclusive hadronic pair production~\cite{sciadv.abq3903,Brandenburg:2024ksp,TOTEM:2024aso,ALICE:2023kgv,STAR:2020dzd,ATLAS:2022uef}, precise measurements of the Drell-S${\rm \ddot{o}ding}$ process offer a new handle to quantify decay-induced effects in nuclear tomography and EESI via comparisons with $\rho^0$ photoproduction. A precise characterization of the Drell-S${\rm \ddot{o}ding}$ process is also important for disentangling excited $\rho^0$ states, e.g., $\rho(1450)$ and $\rho(1700)$, in exclusive $\pi^+\pi^-$ production~\cite{ALICE:2020ugp}.

In this Letter, we report a comprehensive analysis for a series of simultaneous measurements of exclusive $\pi^{+}\pi^{-}$ production. For the first time, (\romannumeral1) the $|t|$ dependence of the Drell-S${\rm \ddot{o}ding}$ process cross section is precisely measured; (\romannumeral2) EESI for the Drell-S${\rm \ddot{o}ding}$ process is measured through ${\rm A_{2\Delta\phi}}$; and (\romannumeral3) the dependence of ${\rm A_{2\Delta\phi}}$ on the pair invariant mass is measured for exclusive $\pi^+\pi^-$ pair production. Finally, the ${\rm A_{2\Delta\phi}}$ data are compared with vector meson dominant (VMD)~\cite{PhysRevD.103.033007,Zha:2018jin} and non-resonance production (NRP) calculations~\cite{pumplin1970diffraction,Zha:2018jin,Wu:2026emu}.

The measurement of exclusive $\pi^{+}\pi^{-}$ production was conducted in Au+Au UPCs at $\sqrt{s_{NN}}=200$ GeV at the Relativistic Heavy Ion Collider (RHIC) using the Solenoidal Tracker at RHIC (STAR)~\cite{STAR:detector}. An inclusive UPC trigger was applied for the pre-selection to the data~\cite{star:trigger}. This trigger relied on rapidity gaps defined by the small beam-beam counter (BBC) tiles on both sides of the detector ($3<\eta<5$)~\cite{star-bbc1,star:bbc2}, low activity in the mid-rapidity in the barrel time-of-flight (TOF) system~\cite{star:tof,Chen:2024aom} (TOF multiplicity between 1 and 7), and the coincidence of neutrons hitting both the east and west zero degree calorimeters (ZDC)~\cite{star:zdc,Xu:2016alq}. The integrated luminosities for the triggered data sets recorded in 2010, 2011, and 2014 are 679, 859, and 1464 $\mu b^{-1}$ with 10\% uncertainty, respectively. The momenta of charged-particle tracks were reconstructed by the time projection chamber (TPC) covering pseudorapidities $\vert \eta \vert<1.0$ for the full azimuthal angle ($0<\phi<2\pi$)~\cite{star:tpc}. 
The TPC also provided specific ionization energy loss (dE/dx) measurements for particle identification, which were used to construct the deviation parameter $n\sigma_\pi$ by comparing the measured dE/dx to the value predicted for a pion.

Further offline cuts on the events were required. Events were required to have a primary vertex reconstructed by a pair of charged-particle tracks with a longitudinal distance of $\vert V_z \vert<100$ cm to the nominal center of STAR. The associated charged-particle tracks were required to have at least 15 fit points and 10 dE/dx points to ensure sufficient momentum and dE/dx resolutions. Charged-particle tracks were required to match to TOF hits in order to mitigate the effect of out-of-time pileup. The distance of closest approach (DCA) for charged-particle tracks was required to be less than 1.5 cm to the primary collision vertex to suppress contamination by secondary decay particles. High purity $\pi^{+}\pi^{-}$ pairs were selected by requiring $\chi^2_{\pi\pi}=n\sigma_{\pi 1}^2+n\sigma_{\pi 2}^2<8$, which was almost fully efficient ($\sim99\%$). 
The azimuthal angle $\Delta\phi$ is defined in the transverse plane as the angle between $\vec{p}_{T1}+\vec{p}_{T2}$ and $\vec{p}_{T1}-\vec{p}_{T2}$, where $\vec{p}_{T1}$ and $\vec{p}_{T2}$ are the transverse momenta of the
$\pi^+$ and $\pi^-$.
Over one million exclusive $\pi^{+}\pi^{-}$ pairs were selected in the invariant mass region $0.6<M_{\pi^+\pi^-}<0.9$ GeV/${\rm c^2}$, enabling precise differential spectra studies.

A Monte Carlo (MC) sample of photoproduced $\pi^+\pi^-$ pairs was generated. The kinematic distributions were weighted according to the model described in Ref.~\cite{Wu:2026emu}. The vertex ($V_z$) distribution was matched to data. The sample data was passed through the STAR detector response simulator using GEANT3~\cite{Brun:1119728} to evaluate the acceptance and detector efficiency. It was then embedded into randomly recorded STAR events that were tuned to reproduce the ZDC coincidence rate during data taking, ensuring realistic background effects on charged-particle track and vertex reconstruction. To minimize potential biases arising from the Monte Carlo sample's underlying kinematic distributions, a high-dimensional correction was employed to obtain the cross section as
\begin{equation}
    \frac{d^3 \sigma}{d p_T d M d \Delta \phi} =\frac{N\left(p_T, M, \Delta \phi\right)}{\varepsilon_{\text {trig }}{\text{corr}}(p_T, M, \Delta \phi)L\Delta p_T\Delta M\Delta(\Delta \phi)},
\end{equation}
where $\varepsilon_{\text {trig }}$ is the trigger efficiency; $\text{corr}(p_T, M, \Delta \phi)$ represents the $p_T$, $M$, and $\Delta \phi$-dependent acceptance and efficiency correction; and $L$ is the integrated luminosity. Since signals were corrected with the $\Delta \phi$ dependence, the amplitudes of the $\Delta \phi$ modulations at all orders were corrected simultaneously. In addition, the momentum smearing of charged-particle tracks in the TPC smooths the $\Delta \phi$ spectra, especially at low $\pi^{+}\pi^{-}$ $p_T$. We estimated the momentum resolution impact and unfold the measured ${\rm A_{2\Delta\phi}}$ based on the ratio of the reconstructed ${\rm A_{2\Delta\phi}}$ to the original in the polarized Monte Carlo simulation, where the $\Delta\phi$ distribution was reweighted according to the polarization model.

An improved description of the exclusive $\pi^{+}\pi^{-}$ mass distribution was constructed to separate the resonance and non-resonance components from different processes, as expressed in Eq.~\ref{eq:fit}. In this parameterization, $A_\rho$ and $C_\omega$ denote the amplitudes for $\rho^0$ and $\omega$, respectively; $B$ represents the magnitude of the normalized model-calculated Drell-S${\rm \ddot{o}ding}$ mass shape $f_{S\ddot{o}ding}$; $\phi_\omega$ is the mixing phase angle for $\omega$; and $f_{\text {bg}}$ is a linear polynomial that accounts for any residual non-photoproduction background. We apply the relativistic Breit-Wigner (rBW) resonance for $\rho^0$ and $\omega$, and a phenomenological non-resonance by the NRP calculation~\cite{Wu:2026emu}. The momentum-dependent mass widths are expressed as~\cite{Alvensleben:1971hz}
\begin{equation}
\begin{aligned}
&\Gamma_\rho=\Gamma_0 \frac{M_\rho}{M_{\pi \pi}}\left(\frac{M_{\pi \pi}^2-4 m_\pi^2}{M_\rho^2-4 m_\pi^2}\right)^{3 / 2},\\
&\Gamma_{\omega \rightarrow \pi \pi}=\operatorname{Br}(\omega \rightarrow \pi \pi) \Gamma_0 \frac{M_\omega}{M_{\pi \pi}}\left(\frac{M_{\pi \pi}^2-4 m_\pi^2}{M_\omega^2-4 m_\pi^2}\right)^{3 / 2},
\end{aligned}
\end{equation}
with $\operatorname{Br}(\omega \rightarrow \pi \pi)=0.0153^{+0.0011}_{-0.0013}$~\cite{ParticleDataGroup:2014cgo}, where $\Gamma_0$ is the pole width for each meson. 
The vector meson photoproduction amplitude is modeled as the photon energy distribution and scattering amplitude for the virtual vector meson and nucleus. Due to the correlation of the photon energy and vector meson mass, $E_{\gamma}=M/2\times exp(\pm y)$, the photon energy distribution slightly modify the rBW mass spectra. This correction, $f_{fluxcorr}$, is considered using VMD model calculation~\cite{PhysRevD.103.033007,Zha:2018jin}. The $\gamma\gamma \rightarrow \mu^+ \mu^-$ process constitutes a background, as muons cannot be distinguished from pions due to the absence of a start time $t_0$ measurement for TOF. Dimuon production, $f_{\mu\mu}$, is calculated using lowest-order QED~\cite{Li:2023yjt,Zha:2018tlq,Zha:2021jhf}, which reproduces experimental data~\cite{STAR:2018ldd,STAR:2019wlg}.

\begin{equation}
\begin{aligned}
\frac{d \sigma}{d M_{\pi^{+} \pi^{-}}} &= \left| A_\rho \frac{\sqrt{M_{\pi \pi} M_{\rho} \Gamma_\rho}}{M_{\pi \pi}^2-M_\rho^2+i M_\rho \Gamma_\rho} f_{fluxcorr} + B f_{S\ddot{o}ding} \right. \\ 
&\left. + C_\omega e^{i \phi_\omega} \frac{\sqrt{M_{\pi \pi} M_\omega \Gamma_{\omega \rightarrow \pi \pi}}}{M_{\pi \pi}^2-M_\omega^2+i M_\omega \Gamma_\omega} \right|^2 + f_{\mu\mu} + f_{\text {bg}}.
\end{aligned}
\label{eq:fit}
\end{equation}

\begin{figure}[h]
  \centering
  \begin{minipage}{0.495\textwidth}
    \includegraphics[width=\textwidth]{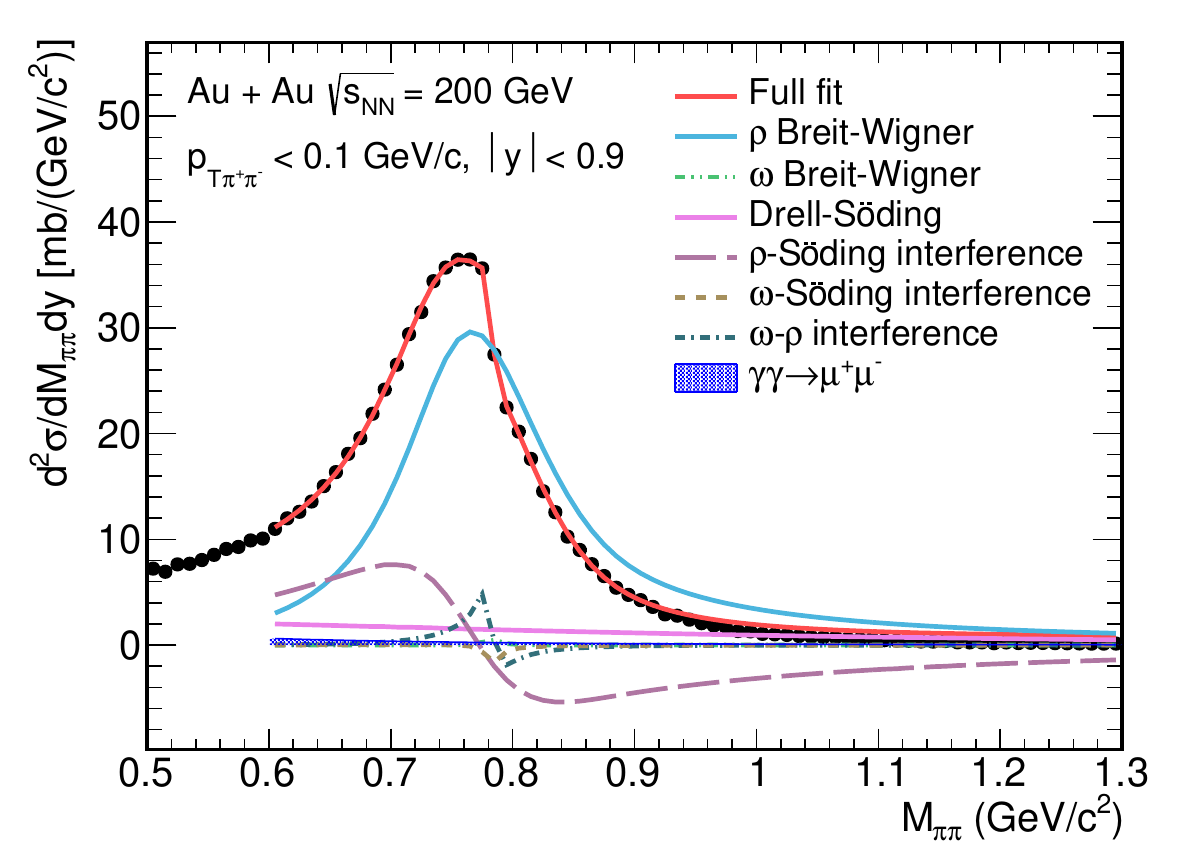}
  \end{minipage}
   \caption{The differential cross section $d^2\sigma/dM_{\pi\pi}dy$ of exclusive $\pi^{+}\pi^{-}$ production for pair $p_T<0.1$ GeV/c and $|y|<0.9$ in Au+Au UPCs at $\sqrt{s_{NN}} = 200$~GeV, together with an illustration of the fitting algorithm to separate the $\rho^{0}$ and Drell-S${\rm \ddot{o}ding}$ components. The black markers show the data, and the lilac line show the Drell-S${\rm \ddot{o}ding}$ term. The interference terms for $\rho$-Söding (purple), $\omega$-Söding (brown), and $\rho-\omega$ (dark green) are shown. $\mu^+\mu^-$ pair production is represented by the blue band.}
  \label{fig:2}
\end{figure}

\begin{figure}[h]
  \centering
  \begin{minipage}{0.48\textwidth}
    \includegraphics[width=\textwidth]{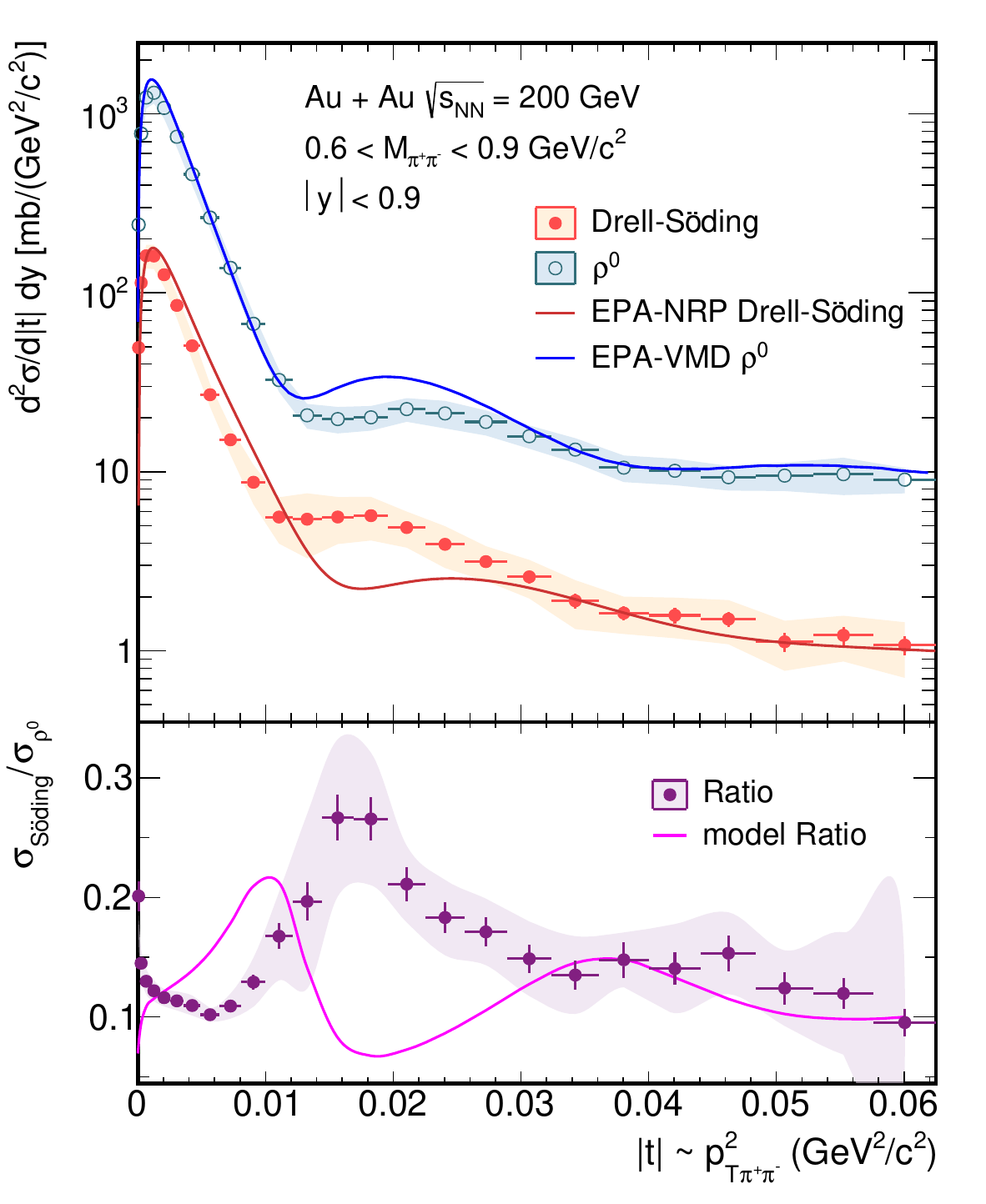}
  \end{minipage}
   \caption{(top) The differential cross section $d^{2}\sigma/d|t|dy$ of exclusive $\pi^{+}\pi^{-}$ production for pair $0.6<M<0.9$ GeV/${\rm c^2}$ and $|y|<0.9$ in Au+Au UPCs at $\sqrt{s_{NN}} = 200$~GeV. The blue and red markers (lines) represent the measured (calculated) $\rho^0$ photoproduction~\cite{Zha:2018jin} and Drell-S${\rm \ddot{o}ding}$ process~\cite{pumplin1970diffraction,bolz2015photoproduction,Wu:2026emu}, respectively. EPA denotes the equivalent photon approximation. (bottom) The measured (calculated) ratio of the Drell-S${\rm \ddot{o}ding}$ to $\rho^0$ photoproduction cross sections.} 
  \label{fig:3}
\end{figure}

\begin{figure*}[ht]
  \centering
  \begin{minipage}{0.96\textwidth}
    \includegraphics[width=\textwidth]{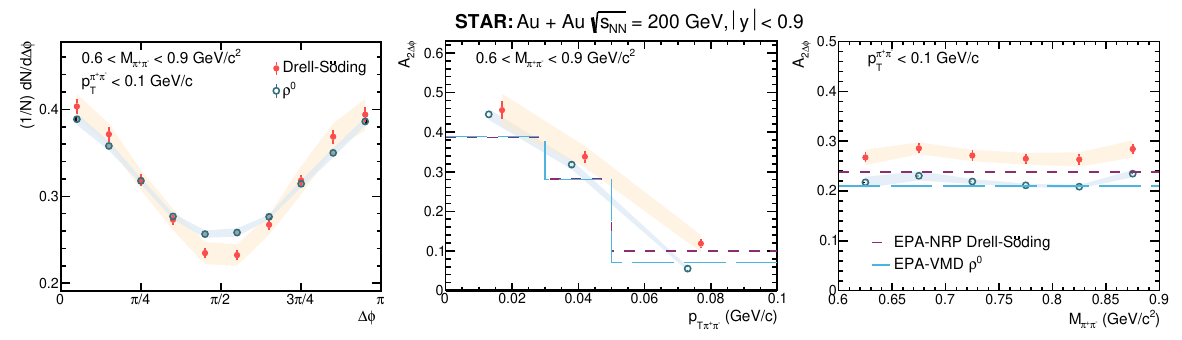}
  \end{minipage}
   \caption{The $\Delta\phi$ distributions normalized to unit integral (left); second-order modulation amplitude $A_{2\Delta\phi}$ versus pair transverse momentum $p_T$ (middle) and invariant mass $M_{\pi^+\pi^-}$ (right) in Au+Au UPC at $\sqrt{s_{NN}}=200$ GeV. The Drell-S${\rm \ddot{o}ding}$ process and $\rho^0$ photoproduction data points are shown as red and teal points. Statistical uncertainties and systematic uncertainties are represented by the error bars and boxes (middle), or bands (left and right) respectively. The model calculations are denoted as dashed lines. In the middle panel, the data and model calculations use the same binning in $p_T$. For clarity, the $\rho^0$ and Drell–S${\rm \ddot{o}ding}$ data points are slightly shifted horizontally relative to the bin centers to avoid overlap.}
  \label{fig:4}
\end{figure*}

Systematic uncertainties are classified into correlated and uncorrelated components among different data sets. Correlated uncertainties were estimated to be 2\% for $\rho^0$ and 5\% for Drell-Söding, arising mainly from the fitting procedure, including choices of fitting range and the modeling of the Drell-S${\rm \ddot{o}ding}$ mass shape. The fitting-range dependence was evaluated by varying the low mass boundary from 0.6 GeV/${\rm c^2}$ to 0.57 GeV/${\rm c^2}$ and 0.65 GeV/${\rm c^2}$.
An alternative parametrization of the non-resonant amplitude, $A_{nr}= f_{nr}/(m_{\pi\pi}^2-4m_\pi^2+\Lambda_{nr}^2)^{\delta_{nr}}$, following Ref.~\cite{H1:2020lzc}, was used to assess the model dependence of the fit to the data, where $f_{n r}$, $\Lambda_{n r}$, and $\delta_{n r}$ are free parameters. 
Uncorrelated uncertainties were evaluated through variations in topological selection criteria and adjustments to the MC model in acceptance and efficiency extractions. Uncorrelated uncertainties are about 5\% for $\rho^0$ photoproduction and 8\% for Drell-S${\rm \ddot{o}ding}$. An external 10\% uncertainty in integrated luminosity determination and a 3\% per-track reconstruction uncertainty were included. 
Results from different data sets are statistically consistent within uncertainties and are combined using a weighted average based on the quadratic sum of statistical and uncorrelated systematic uncertainties. The statistical and uncorrelated systematic uncertainties were treated as independent and were propagated to the combined result by summing their contributions in quadrature with the corresponding weights.
The correlated systematic uncertainties, assumed to be fully correlated across data sets, were combined by taking their weighted average.
The correlations between $\rho^0$ photoproduction and Drell-S${\rm \ddot{o}ding}$ were accounted for through the covariance matrix in the fit.
The total systematic uncertainty was obtained by adding the uncorrelated and correlated uncertainties in quadrature. 
The systematic uncertainty sources and evaluation procedure are the same for the extracted $A_{2\Delta\phi}$. For the difference between the $A_{2\Delta\phi}$ of $\rho^0$ and Drell-S${\rm \ddot{o}ding}$, the uncertainty was evaluated directly on the difference, such that the correlations between them are naturally included.

Figure~\ref{fig:2} illustrates the differential cross section $d^2\sigma/dM_{\pi\pi}dy$ as a function of the invariant mass $M_{\pi\pi}$.
The data points represent the measured cross section, while the red curve shows the full fit incorporating various contributing components.
Although the Drell-S${\rm \ddot{o}ding}$ process constitutes a relatively minor contribution to exclusive $\pi^+\pi^-$ production, its interference with the $\rho^0$ production crucially shapes the observed mass spectrum. The mass shift and the asymmetric structure of the mass peak are both governed by this interference thus enabling a precise extraction of the Drell-Söding contribution. Mass spectra were fitted independently in each ($p_T$, $\Delta \phi$) bin to extract the contributions of the different processes and evaluate the corresponding cross sections.
Amplitudes of the second order cosine modulation ${\rm A_{2\Delta\phi}}$ was reconstructed by a weighted sum ${\rm A_{2\Delta\phi}}=\sum_{i=1}^n 2N_i{\rm cos(2\Delta \phi_i)}$, 
where $N_i$ denotes the yield in each $\Delta \phi_i$ bin normalized by the total yield in the corresponding $p_T$ interval.
The fit quality is quantified using the reduced chi-square ($\chi^2/\text{ndf}$), which ranges from 50 to 100 per 65 degrees of freedom, reflecting consistent modeling of the spectral features.

Figure~\ref{fig:3} presents the differential cross-section $d^2\sigma/d|t| dy$ for the Drell-S${\rm \ddot{o}ding}$ process and $\rho^0$ photoproduction, along with their ratio as a function of $|t| \approx p_{T\pi^{+}\pi^{-}}^{2}$. The $\rho^0$ results are consistent with the previous measurement~\cite{STAR:2017enh}.
Compared to $\rho^0$ photoproduction, the Drell-S${\rm \ddot{o}ding}$ process exhibits a steeper decline with increasing $|t|$ with its diffractive minima shifted to lower $|t|$, reflecting a softer production mechanism and differences in the underlying photon hadronic structure.
The measured ratio reveals a pronounced bump at low $|t|$, suggesting that the virtual $\pi^+\pi^-$ configuration is spatially broader than the $\rho^0$ state. Theoretical calculations do not reproduce this trend, possibly due to missing dipole-size effects that are not accounted for in the model.

Figure~\ref{fig:4} shows the azimuthal anisotropy (left), quantified by the second-order cosine modulation amplitude, ${\rm A_{2\Delta\phi}}$, for the Drell-S${\rm \ddot{o}ding}$ process and $\rho^0$ photoproduction as functions of the pair transverse momentum $p_T$ (middle) and the invariant mass $M_{\pi^+\pi^-}$ (right) within rapidity $|y|<0.9$. 
The amplitudes ${\rm A_{2\Delta\phi}}$ were then extracted from the fitted yields in distinct $\Delta\phi$ or $p_T$--$\Delta\phi$ intervals.
To ensure sufficient statistical precision for resolving subtle features in the mass distribution, the $p_T$ range is divided into three bins: $0 < p_T < 0.03~\mathrm{GeV}/c$, $0.03 < p_T < 0.05~\mathrm{GeV}/c$, and $0.05 < p_T < 0.1~\mathrm{GeV}/c$. The resulting $\Delta\phi$ distributions are then used to extract the $\cos(2\Delta\phi)$ modulation amplitude. 
The ${\rm A_{2\Delta\phi}}$ of the Drell-S${\rm \ddot{o}ding}$ process exhibits a similar $p_T$ dependence as $\rho^0$ photoproduction and has an enhancement in transverse momentum in the range $0.05<p_T<0.1$ GeV/c. 
The right panel illustrates the $M_{\pi^+\pi^-}$ dependence of ${\rm A_{2\Delta\phi}}$. 
No significant $M_{\pi^+\pi^-}$ dependence of ${\rm A_{2\Delta\phi}}$ is found, indicating no strong photon energy dependence in EESI. Notably, the extracted ${\rm A_{2\Delta\phi}}$ for the Drell-S${\rm \ddot{o}ding}$ process is systematically higher than that for $\rho^0$ photoproduction.
For $0 < p_T < 0.1~\mathrm{GeV}/c$, $|y| < 0.9$, and $0.6 < M_{\pi^+\pi^-} < 0.9~\mathrm{GeV}/c^2$, the extracted ${\rm A_{2\Delta\phi}}$ values are $0.273 \pm 0.010 \pm 0.013$ (Drell-S${\rm \ddot{o}ding}$) and $0.219 \pm 0.003 \pm 0.005$ ($\rho^0$ photoproduction). The first quoted errors are statistical uncertainties, while the second are systematic uncertainties. The resulting difference is $0.054 \pm 0.010 \pm 0.015$, corresponds to a significance of 3.0~$\sigma$ when statistical and systematic uncertainties are included. 

The predictions of the EPA-VMD and NRP model are also compared with the data and found to underestimate $A_{2\Delta\phi}$ at low transverse momentum, $0<p_T<0.05$ GeV/c. 
The results show no evidence, within the present uncertainties, for a loss of coherence in scattering off a nucleus, nor for destructive interference induced by the difference between the $\pi^+p$ and $\pi^-p$ scattering amplitudes in the Drell-S${\rm \ddot{o}ding}$ process.

In conclusion, we report the observation of EESI in the Drell-S${\rm \ddot{o}ding}$ process and the pair $p_T$ and $M$ dependence of the differential cross section for the Drell-S${\rm \ddot{o}ding}$ process and $\rho^0$ photoproduction in Au+Au UPC at $\sqrt{s_{NN}}=200$ GeV at RHIC. A model-based Drell-S${\rm \ddot{o}ding}$ shape is employed to separate different processes involved in exclusive $\pi^+\pi^-$ production. 
The observation of EESI effects demonstrates that the non-resonant $\pi^+\pi^-$ pairs are entangled in the Drell-S${\rm \ddot{o}ding}$ process. The amplitude, ${\rm A_{2\Delta\phi}}$, for the Drell-S${\rm \ddot{o}ding}$ process and $\rho^0$ photoproduction are found to have similar decreasing trends along the pair $p_T$. Compared to the $\rho^0$, an enhancement is observed in the Drell-S${\rm \ddot{o}ding}$ process in pair $p_T$ in the range $0.05<p_T<0.1$ GeV/c. The measured ${\rm A_{2\Delta\phi}}$ for these processes is found to have no obvious pair mass $M_{\pi^+\pi^-}$ dependence, and the amplitude ${\rm A_{2\Delta\phi}}$ for Drell-S${\rm \ddot{o}ding}$ is 3.0 $\sigma$ higher than for $\rho^0$ photoproduction over the observed mass regions. Moreover, the observed Drell-S${\rm \ddot{o}ding}$ $|t|$ spectrum is softer when compared to $\rho^0$ photoproduction. These results advance the exploration of quantum interference phenomena in photon–nucleus interactions, providing a basis for future investigations in ultraperipheral heavy-ion collisions.


\textit{Acknowledgments}—We thank the RHIC Operations Group and SDCC at BNL, the NERSC Center at LBNL, and the Open Science Grid consortium for providing resources and support.  This work was supported in part by the Office of Nuclear Physics within the U.S. DOE Office of Science, the U.S. National Science Foundation, National Natural Science Foundation of China, Chinese Academy of Science, the Ministry of Science and Technology of China and the Chinese Ministry of Education, NSTC Taipei, the National Research Foundation of Korea, Czech Science Foundation and Ministry of Education, Youth and Sports of the Czech Republic, Hungarian National Research, Development and Innovation Office, New National Excellency Programme of the Hungarian Ministry of Human Capacities, Department of Atomic Energy and Department of Science and Technology of the Government of India, the National Science Centre and WUT ID-UB of Poland, German Bundesministerium f\"ur Bildung, Wissenschaft, Forschung and Technologie (BMBF), Helmholtz Association, Ministry of Education, Culture, Sports, Science, and Technology (MEXT), and Japan Society for the Promotion of Science (JSPS).

\bibliography{referencenew}

\end{document}